\documentclass[11pt]{article}%
\usepackage{amsmath}%
\usepackage{amsfonts}%
\usepackage{amssymb}%
\usepackage{graphicx}

\begin{document}

\title{{\huge Soft Photon Spectrum in Orthopositronium and Vector Quarkonium Decays}\\\ \linebreak \\\ }
\author{J. Pestieau, C. Smith
\and \linebreak \\\textsl{Institut de Physique Th\'{e}orique, Universit\'{e} catholique de
Louvain,}\linebreak \\\ \ \textsl{\linebreak Chemin du Cyclotron, 2, B-1348, Louvain-la-Neuve}}
\date{November 21, 2001}
\maketitle

\begin{abstract}
\textsf{QED gauge invariance, when combined with analyticity, leads to
constraints on the low energy end of the emitted photon spectra. This is known
as Low's theorem. It is shown that the Ore-Powell result, as well as further
developments for the orthopositronium differential decay rate, are in
contradiction with Low's theorem, i.e. that their predicted soft photon
spectra are incorrect.}

\textsf{A solution to this problem is presented. The implications for the
orthopositronium lifetime puzzle, the charmonium }$\rho-\pi$\textsf{\ puzzle,
the prompt photon spectrum in inclusive quarkonium decays and the extraction
of }$\alpha_{S}$\textsf{\ from quarkonium annihilation rates are briefly commented.}

\pagebreak 

\end{abstract}

The single most important concept in contemporary physics is probably gauge
invariance. Aside from that, a basic postulate of Quantum Field Theory, and
therefore also of QED, is the analyticity of probability amplitudes as
functions of their kinematical variables. In the present note, we will
investigate the consequences of those two principles, analyticity and gauge
invariance, for orthopositronium decay into three photons, and quarkonium
decay into three photons, two gluons plus a photon or three gluons.

The consequences of gauge invariance are well-known: one gets very though
constraints on the structure of amplitudes in the form of Ward-Takahashi
Identities. For instance, any amplitude involving an external photon,
$\mathcal{M}=\varepsilon_{\mu}\mathcal{M}^{\mu}$, must verify the Ward
Identity $k_{\mu}\mathcal{M}^{\mu}=0$, with $k_{\mu}$ the photon momentum and
$\varepsilon_{\mu}$ its polarization vector. In addition to gauge invariance,
since any probability amplitude is an analytical function, $\mathcal{M}^{\mu
}\left(  k,...\right)  $ admit a Laurent expansion in each of its variables.
It was F.E. Low who, in the fifties \cite{Low}, first realized that the Ward
identity restricts the form of the first two terms of the Laurent expansion in
the external photon energy.

Low's theorem is a model-independent result, valid to all orders: for a
complete amplitude, the soft-photon limit only depends on the quantum numbers
of the external particles, and not on the details of the intermediate
subprocesses. At the level of observables, the low-energy end of the photon
spectrum is obtained by combining the amplitude behavior with that of the
phase-space. The most characteristic spectra are\medskip

-- \textit{Charged particles and photons in external states}. The well-known
bremsstrahlung emissions lead to an amplitude in $1/\omega$ for $\omega
\rightarrow0$ ($\omega$ is the energy of one of the emitted photons). At the
decay rate or cross section level, the IR divergent amplitude generates
characteristic IR divergent spectra.\medskip

-- \textit{Only neutral bosons, including photons in external state.} The
amplitude is in $\omega$ for $\omega\rightarrow0$ ($\omega$ is again the
energy of one of the photon). This statement is much stronger than what is
sometimes thought of for a non-bremsstrahlung process. \textit{On general
ground, it shows that feeling confident with an IR safe computation is
theoretically incorrect}. IR safety at the cross section or decay rate level
is definitely not sufficient. When one construct a model designed to describe
some processes among neutral bosons and photons, one must ensure that the
amplitude \textbf{vanishes} in the soft-photon limit.\medskip

The present note is organized as follow. First, we show that the lowest order
result of Ore and Powell for orthopositronium \cite{OrePowell}, or vector
quarkonium, decay into $\gamma\gamma\gamma$, is in contradiction with Low's
theorem. Then, we discuss the sources of the problem, and argue that present
theoretical models (and therefore the corrections computed with them) are
incomplete. We then propose (and motivate by comparison with standard
elementary particle processes) a model that respect Low's theorem. Finally, in
the conclusions, we shortly comment about the consequences for
orthopositronium lifetime puzzle and heavy quarkonia puzzles.

\section{Contradiction between Low's theorem\newline and Ore-Powell
differential decay rate}

We will concentrate on the orthopositronium decay into three photons. The
discussion also applies to vector quarkonium ($J/\psi$, $\Upsilon$,...).

From the requirement of gauge invariance, and because of the quantum numbers
of the initial and final particles involved (neutral self-conjugate bosons),
Low's theorem predicts that \textit{the decay amplitude must vanish linearly
when the energy of one of the photons is going to zero}
\begin{equation}
\mathcal{M}\left(  o\text{-}Ps\rightarrow\gamma\gamma\gamma\right)
\overset{\omega\sim0}{\sim}\mathcal{O}\left(  \omega\right) \label{LowAmps}%
\end{equation}
with $\omega$ the energy of one of the photon. The squared modulus of the
amplitude therefore behaves as $\mathcal{O}\left(  \omega^{2}\right)  $ for
small photon energy. Aside from that, the three-photon phase-space alone (i.e.
with a constant decay amplitude) gives a differential rate as
\begin{equation}
\left.  \frac{d\Gamma\left(  o\text{-}Ps\rightarrow\gamma\gamma\gamma\right)
}{dx}\right|  _{Phase-Space}\sim x\label{PS}%
\end{equation}
with $x=2\omega/M$ the reduced photon energy, $M$ the orthopositronium mass.
Combining the amplitude with the phase-space, one finds that the low-energy
end of the photon spectrum \textbf{must} behave as
\[
\frac{d\Gamma\left(  o\text{-}Ps\rightarrow\gamma\gamma\gamma\right)  }%
{dx}\overset{x\sim0}{\sim}x^{3}%
\]

No matter the model used to compute the orthopositronium decay rate, the
differential rate must exhibit this $x^{3}$ behavior for small photon
energies. As we have already pointed out in \cite{PreviousWork}, the lowest
order decay amplitude, as found by Ore and Powell, leads to a differential
decay rate in contradiction with the analytical requirement of Low's theorem.
Their model is based on the formula \cite{History}:
\begin{equation}
\Gamma\left(  o\text{-}Ps\rightarrow\gamma\gamma\gamma\right)  =\frac{1}%
{3}\left|  \psi\left(  0\right)  \right|  ^{2}\left(  4v_{rel}\sigma\left(
e^{-}e^{+}\rightarrow\gamma\gamma\gamma\right)  \right)  _{v_{rel}%
\rightarrow0}\label{Pirenne}%
\end{equation}
with $v_{rel}$ the relative velocity of the $e^{+}e^{-}$ in their
center-of-mass frame, and $\psi\left(  0\right)  $ the positronium
Schr\"{o}dinger wavefunction at zero separation ($m$ is the electron mass)
\begin{equation}
\left|  \psi\left(  0\right)  \right|  ^{2}=\frac{\alpha^{3}m^{3}}{8\pi
}\label{SchroWF}%
\end{equation}
This formula states that in first approximation, the positronium decay rate
can be computed from the static limit of the scattering cross section
$e^{+}e^{-}\rightarrow\gamma\gamma\gamma$. Equivalently, it is found from the
squared modulus amplitude for an $e^{+}e^{-}$ pair at rest into $\gamma
\gamma\gamma$. Summed over photon polarizations, this is easily shown to be
\cite{OrePowell}
\[
\sum_{polarizations}\left|  \mathcal{M}\left(  \left(  e^{+}e^{-}\right)
_{v_{rel}=0}\rightarrow\gamma\gamma\gamma\right)  \right|  ^{2}=\frac{\left(
1-x_{1}\right)  ^{2}}{x_{2}^{2}x_{3}^{2}}+\frac{\left(  1-x_{2}\right)  ^{2}%
}{x_{1}^{2}x_{3}^{2}}+\frac{\left(  1-x_{3}\right)  ^{2}}{x_{1}^{2}x_{2}^{2}}%
\]
which behaves as a constant when one of the $x_{i}$ is vanishing (as can be
seen by using energy-momentum conservation $x_{1}+x_{2}+x_{3}=2$), while it
should vanish as $x_{i}^{2}$. In turn, the well-known differential rate
inherits an incorrect analytical behavior
\[
\frac{d\Gamma\left(  o\text{-}Ps\rightarrow\gamma\gamma\gamma\right)  }%
{dx_{1}}=\frac{2\alpha^{6}m}{9\pi}\Omega\left(  x_{1}\right)
\]
where the spectrum function is
\begin{align}
\Omega\left(  x_{1}\right)   & =\int_{1-x_{1}}^{1}dx_{2}\left|  \mathcal{M}%
\left(  \left(  e^{+}e^{-}\right)  _{v_{rel}=0}\rightarrow\gamma\gamma
\gamma\right)  \right|  _{x_{3}=2-x_{1}-x_{2}}^{2}\label{SpectrumFunc}\\
& =\frac{2\left(  2-x_{1}\right)  }{x_{1}}+\frac{2\left(  1-x_{1}\right)
x_{1}}{\left(  2-x_{1}\right)  ^{2}}+4\left[  \frac{\left(  1-x_{1}\right)
}{x_{1}^{2}}-\frac{\left(  1-x_{1}\right)  ^{2}}{\left(  2-x_{1}\right)  ^{3}%
}\right]  \ln\left(  1-x_{1}\right) \nonumber\\
& =\frac{5}{3}x_{1}+\mathcal{O}\left(  x_{1}^{2}\right)  \text{\ near }%
x_{1}=0\nonumber
\end{align}
In the Ore-Powell model, the photon energy spectrum vanishes only linearly
near zero, instead of the required $\Omega\left(  x_{1}\right)  =\mathcal{O}%
\left(  x_{1}^{3}\right)  $.\medskip

\textbf{Remarks}

For completeness, recall that it is this differential rate that gives the
total width
\[
\Gamma\left(  o\text{-}Ps\rightarrow\gamma\gamma\gamma\right)  =\frac{2\left(
\pi^{2}-9\right)  }{9\pi}\alpha^{6}m\text{\ \ \ \ since }\int_{0}^{1}%
dx_{1}\Omega\left(  x_{1}\right)  =\pi^{2}-9
\]
Up to color factors, wavefunctions, coupling constants, the present analysis
can be repeated for the quarkonia decay modes \cite{QuarkoniaAnnihRate}
\begin{align*}
\frac{d\Gamma\left(  V\rightarrow\gamma\gamma\gamma\right)  }{dx_{1}}  &
=\frac{64}{3}e_{Q}^{6}\alpha^{3}\frac{\left|  \phi_{0}\right|  ^{2}}{M^{2}%
}\Omega\left(  x_{1}\right)  \\
\frac{d\Gamma\left(  V\rightarrow ggg\right)  }{dx_{1}}  & =\frac{160}%
{81}\alpha_{S}^{3}\frac{\left|  \phi_{0}\right|  ^{2}}{M^{2}}\Omega\left(
x_{1}\right)  \\
\frac{d\Gamma\left(  V\rightarrow gg\gamma\right)  }{dx_{1}}  & =\frac{128}%
{9}e_{Q}^{2}\alpha\alpha_{S}^{2}\frac{\left|  \phi_{0}\right|  ^{2}}{M^{2}%
}\Omega\left(  x_{1}\right)
\end{align*}
where $V$ is the $1^{-\,-}$ vector bound state made of the $Q\bar{Q}$ pair,
$M$ is the mass of $V$, $\phi_{0}$ the (unknown) quarkonium configuration
space wavefunction at zero separation and $e_{Q}$ the heavy quark electric
charge in units of the electron one. All these decay spectra do violate the
basic requirement of analyticity.

\section{Contradiction between Low's theorem\newline and current positronium
decay models}

The model used by Ore-Powell, based on the factorized formula (\ref{Pirenne}),
may seem a bit naive, especially in view of the enormous amount of work done
by various groups (see for example \cite{PreviousWork}, \cite{RadiativeCorr},
\cite{NRQED}, \cite{LargeRelCorr}, and references quoted there). Nevertheless,
it is very illustrative of the current approaches in its treatment of
intermediate states. Indeed, models derived from Bethe-Salpeter analyses, or
from QED non-relativistic effective theory (NRQED, \cite{NRQED}), always
connect the process of annihilation of bound charged particles to that of
scattering of real, asymptotic charged particles. The difficulty with such
approaches is thereby apparent: asymptotic and bound charged particles have
drastically different radiation properties: the former exhibit
bremsstrahlung-type radiations, while the later do not radiate zero energy
photons (for very low-energy photons, a positronium state is just a neutral,
self-conjugate boson, hence it does not radiate in that limit).

Bremsstrahlung radiations typically lead to a Laurent expansion for the
amplitude as
\[
\mathcal{M}^{\mu}\left(  \omega,...\right)  =\mathcal{O}\left(  1/\omega
\right)  +\mathcal{O}\left(  1\right)  +\mathcal{O}\left(  \omega\right)
\]
What Low's theorem state is that both the terms of $\mathcal{O}\left(
1/\omega\right)  $ \textit{and} $\mathcal{O}\left(  1\right)  $ must disappear
\cite{Low}. While the cancellation of $\mathcal{O}\left(  1/\omega\right)  $
terms is automatic from selection rules, that of $\mathcal{O}\left(  1\right)
$ terms is much more delicate, requiring a non-perturbative treatment of the
binding energy. Typically, the $\mathcal{O}\left(  \omega\right)  $ term is of
the form \cite{NewLetter}
\begin{equation}
\mathcal{M}^{\mu}\left(  \omega,...\right)  \overset{\omega\rightarrow0}{\sim
}\omega\left(  \frac{M^{2}}{M^{2}-4m^{2}}+\text{regular terms as }%
M\rightarrow2m\right) \label{BindingEffects}%
\end{equation}
Obviously, a perturbative expansion in the binding energy $M-2m$ is
mathematically inconsistent with the soft-photon expansion: if the limit
$M\rightarrow2m$ is taken before $\omega\rightarrow0$, spurious $\mathcal{O}%
\left(  1\right)  $ terms arise. Because the basis of current computations is
a perturbative expansion in the binding energy $M-2m$, computed as
relativistic and radiative corrections to the Ore-Powell result, one can
expect that spurious radiations affect the corrections presented in the literature.

As a comment, notice that the present considerations apply to amplitudes.\ For
orthopositronium, any term less singular than $\mathcal{O}\left(
1/\sqrt{\omega}\right)  $ would lead to an IR finite decay rate, because of
(\ref{PS}). In other words, unphysical terms can lead to IR finite
contributions to the decay rate. Computations presented in the literature
sometimes explicitly exhibit such a bad behavior \cite{LargeRelCorr2}.

\ \newline 

To illustrate how the Low's theorem is implemented in the presence of charged
intermediate states, let us give an example: $K_{S}^{0}\rightarrow e^{+}%
e^{-}\gamma$. The process is modeled by a charged pion loop. The resulting
amplitude can be found in many places (see for example \cite{PreviousWork},
\cite{KaonPion}, \cite{DispRel}), and it behaves exactly as predicted by Low's
theorem (i.e. as $\omega^{3}$ near $\omega=0$), even if in this case
$M_{K}>2m_{\pi}$, i.e. the intermediate charged pion pair can be on-shell.

Now, let us imagine that one is willing to compute the decay rate for
$K_{S}^{0}\rightarrow e^{+}e^{-}\gamma$ by assuming that intermediate on-shell
$\pi^{+}\pi^{-}$ dominates. The decay process is then factorized as $K_{S}%
^{0}\rightarrow\pi^{+}\pi^{-}$ times $\pi^{+}\pi^{-}\rightarrow e^{+}%
e^{-}\gamma$. This is exactly the approximation done to get the Ore-Powell
result: $o$-$Ps\rightarrow e^{+}e^{-}$ times $e^{+}e^{-}\rightarrow
\gamma\gamma\gamma$. However, the soft-photon spectrum of the factorized
approximation is completely wrong, being in contradiction with Low's theorem.

The approximation done was too stringent. To get the correct answer, one must
also consider processes like $K_{S}^{0}\rightarrow\pi^{+}\pi^{-}\gamma$ times
$\pi^{+}\pi^{-}\left(  \gamma\right)  \rightarrow e^{+}e^{-}\left(
\gamma\right)  $ (where the photon is disconnected). These bremsstrahlung
processes interfere destructively with the previous ones, giving a finite
\textbf{vanishing} complete amplitude in the soft-photon limit. In the
framework of dispersion relations \cite{DispRel}, this is a simple application
of the Cutkosky rule to get the absorptive part of an amplitude. Similarly, we
state that the reason why the Ore-Powell result fails to exhibit a correct
soft-photon spectrum is because some contributions to the amplitude are missed
(like $o$-$Ps\rightarrow e^{+}e^{-}\gamma$ times $e^{+}e^{-}\left(
\gamma\right)  \rightarrow\gamma\gamma\left(  \gamma\right)  $, see
\cite{PreviousWork}, \cite{NewLetter}).

Higher order corrections to the Ore-Powell result \cite{RadiativeCorr}
\cite{LargeRelCorr} are similarly incomplete. Trying to take binding energy
effects into account with models based on any kind of factorization
$o$-$Ps\rightarrow e^{+}e^{-}$ times $e^{+}e^{-}\rightarrow\gamma\gamma\gamma$
is hopeless in view of Cutkosky rules. The fact that the on-shell intermediate
state method ''works'' (i.e. is not IR divergent at the decay rate level) for
orthopositronium is not a valid argument, since the amplitudes produced by
that method fails to fulfill a basic requirement of QED, namely Low's theorem.

\section{Conclusions and Perspectives}

From a theoretical perspective, as shown in \cite{PreviousWork}, the
introduction of additional contributions (i.e. process like $o$-$Ps\rightarrow
e^{+}e^{-}\gamma$ times $e^{+}e^{-}\left(  \gamma\right)  \rightarrow
\gamma\gamma\left(  \gamma\right)  $) to the positronium decay amplitude is
unavoidable if one is willing to fulfill the basic requirement of Low's
theorem. As we have explained, the $\omega^{3}$ low-energy spectrum is a
consequence of the properties of the positronium or quarkonium ''as seen from
far away'', i.e. as a neutral point-like self-conjugate bosonic particle. Such
a particle does not radiate zero-energy photons, and the resulting photon
spectrum must exhibit a $\omega^{3}$ shape near $\omega=0$. This is equally
true for $J/\psi$, $\Upsilon$..., and our solution is naturally extended to
quarkonium theory.

On the practical side, it appears quite obvious that the violation of Low's
theorem is very small in positronium decay. In other words, the missing
contributions are subleading. As discussed in \cite{PreviousWork}, we can
expect them to introduce corrections of the order of the binding energy
$E_{B}=M-2m$, i.e. $\alpha^{2}$, or beyond. Nevertheless, even if very small,
those corrections are relevant to the current theoretical considerations.
Indeed, it is at the $\alpha^{2}$ level that some discrepancies have been
found among experiments, the so-called \textbf{orthopositronium lifetime
puzzle} \cite{PreviousWork}, \cite{RadiativeCorr}, \cite{LargeRelCorr}. What
we claim is that no definite answer to that puzzle could be given at present.
Indeed, the additional contributions could turn out to be less suppressed than
usually thought, and the current theoretical result for the $\alpha^{2}$
correction is not fully reliable.

This state of affair is to be contrasted to the quarkonium case. There, the
missing contributions could become sizeable since the binding energy is
non-negligible (see (\ref{BindingEffects})). In other words, the relevance of
the Ore-Powell result for quarkonia is doubtful for any precision calculation.

For instance, the \textbf{photon spectrum in inclusive quarkonium decay into
hadrons + photon} \cite{PromptPhoton} will clearly be affected by the missing
contributions. Indeed, the modification of the spectrum needed at low energy
to fulfill Low's theorem will affects the spectrum also at high energy (since
when the photon has its maximum energy, one of the gluon's energy can go to
zero, see the integration ranges in (\ref{SpectrumFunc})).

Also, the missing contributions should be crucial to solve \textbf{the }%
$\rho-\pi$\textbf{\ puzzle} \cite{RhoPiGeneral}, \cite{RhoPi}. The so-called
$14\%$ rule is obtained from the ratio of the leptonic mode of the $J/\psi$
and $\psi\left(  2S\right)  $ (which is essentially the ratio of wavefunctions
at zero separation). There is no missed contribution for the leptonic modes.
On the other hand, at least $12$ additional contributions need to be
considered for three-gluon modes, and those will depend on the binding energy,
which is quite different for $J/\psi$ and $\psi\left(  2S\right)  $. Those
additional contributions, which arise already at the lowest order, are
essential to enforce Low's theorem through their destructive interferences
with the standard factorized Ore-Powell ones.

Finally, the previous remark also shows that \textbf{the extraction of
}$\alpha_{S}$ \cite{QuarkoniaAnnihRate} from quarkonia branching fractions
will be affected, at least partially, by the additional contributions.\newline 

In conclusion, we have shown that the implications of gauge invariance and
analyticity, in the form of constraints on the low-energy end of photon
spectrum are not met by current bound state decay models. Restoring a correct
behavior in that low-energy region could lead to potentially interesting
advances in both QED and QCD bound state description.

\qquad\newline 

{\Large Acknowledgments: }We are very pleased to acknowledge useful
discussions with Gabriel Lopez Castro, Jean-Marc G\'{e}rard, St\'{e}phanie
Trine and Jacques Weyers. C. S. acknowledges financial support from FNRS and
IISN (Belgium).

\end{document}